\documentclass[runningheads,a4paper]{llncs}

\usepackage{amssymb}
\usepackage{graphicx}
\usepackage{todonotes}
\usepackage{url}
\urldef{\mailsa}\path|rasmus@uoregon.edu|
\urldef{\mailsb}\path|dangle@ucar.edu, soren.rasmussen@aggiemail.usu.edu|
\newcommand{\keywords}[1]{\par\addvspace\baselineskip
\noindent\keywordname\enspace\ignorespaces#1}

\usepackage{cite}

\begin{document}

\mainmatter  

\title{Locally-Oriented Programming:\\ A Simple Programming Model for Stencil-Based Computations
       on Multi-Level Distributed Memory Architectures}

\titlerunning{Locally-Oriented Programming}

\author{Craig Rasmussen
\and Matthew Sottile\and Daniel Nagle\and Soren Rasmussen}

\authorrunning{Locally-Oriented Programming}


\institute{University of Oregon, Eugene, Oregon, USA\\
\mailsa
}

%
%

\maketitle

\begin{abstract}

Emerging hybrid accelerator architectures for high performance computing are often suited for the
use of a data-parallel programming model.  Unfortunately, programmers of these
architectures face a steep learning curve that frequently requires learning a new language (e.g.,
OpenCL). Furthermore, the distributed (and
frequently multi-level) nature of the memory organization of clusters of these machines
provides an additional level of complexity.  This paper presents preliminary work
examining how programming with a local orientation can be employed to provide simpler access to
accelerator architectures.  A locally-oriented programming model is especially useful for the
solution of algorithms requiring the application of a stencil or convolution kernel.  In this programming
model, a programmer codes the algorithm by modifying \emph{only a single array element} (called the
local element), but has read-only access to a small sub-array surrounding the local element.
We demonstrate how a locally-oriented programming model can be adopted as
a language extension using source-to-source program transformations.

\end{abstract}

\section{Introduction}

Historically it has been hard to create parallel programs.  The
responsibility (and difficulty) of creating \emph{correct} parallel
programs can be viewed as being spread between the programmer and the
compiler.  Ideally we would like to have parallel languages that make
it easy for the programmer to express correct parallel programs --- and
conversely it should be difficult to express incorrect parallel
programs.  Unfortunately, many current languages and standards place
all of the responsibility on the user.  The best example of this are
programs written using MPI (Message Passing Interface), where the
programmer expresses all parallelism in terms of calls to library
routines and the serial C or Fortran compiler knows nothing about the
parallel semantics of the combined language plus MPI library.

Of course one would hope that over time code complexity goes down as
better languages allow compilers to take on a greater share of the
parallel complexity burden.  Multithreaded code written using OpenMP
can be significantly simpler than the corresponding code in which the
programmer explicitly manages a pool of threads themselves.  This is
somewhat true for distributed memory parallelism using Unified
Parallel C (UPC) and Coarray Fortran.  With these PGAS
(Partitioned Global Address Space) extensions, C and
Fortran compilers are now aware of parallelism and now generate
message passing code that previously had been handled by the MPI
library.  In some instances the compiler is able to perform
optimizations that it is not able to do with a library-based scheme
like MPI \cite{preissl:2011:SC}.


However, in large part these languages are mostly syntactic sugar for
message passing and do not provide a substantial decrease in code
complexity when application performance is a
goal\cite{hasert:2011:EuroMPI}.  While skilled programmers in the HPC
community have become accustomed to the level of complexity of an MPI
program, the problem for programmers is that hardware is changing in
ways that increase the level of on-chip parallelism.  Indeed, the
current generation of machines could be the last to provide
homogeneous multi-core parallelism\cite{JORSjors.aw}.  For ultimate
performance, programmers must envision coding for heterogeneous
architectures with huge new levels of on-node parallelism, at the same
time they account for off-node parallelism.
Since languages have not evolved that allow the compiler to take up
this increase, the complexity for a programmer has necessarily
dramatically increased.  Unfortunately, considering the complexity of
large multiphysics code frameworks, it could take several person years
to retarget an application for \emph{each} new heterogeneous
platform\cite{DBLP:journals/corr/DubeyS13}.

A reasonable solution given today's language limitations is to use MPI for distributed memory
parallelism \emph{plus} OpenMP for on-chip parallelism.  This hybrid approach is taken by two
successful, large-scale AMR multiphysics code frameworks, FLASH and Chombo; currently however,
neither of these codes support hardware accelerators\cite{DBLP:journals/corr/DubeyS13}.
Other choices for expressing on-chip parallelism are OpenCL~\cite{opencl11} and NVIDIA's CUDA.
Unfortunately, achieving high performance using either of these two languages can be a daunting
challenge and there are no guarantees that either language is suitable for future hardware
generations.

In this paper we examine a language-based paradigm that allows the
compiler to take on a larger portion of the code complexity burden.
Anyone programming in OpenCL is aware that explicit loop
structures over array elements in a serial program are removed and
replaced by a kernel program that is run over all of the elements of
the input arrays.  We propose \emph{Locally Orientated Programming
extensions} (LOPe) to the Fortran and potentially C languages that
formally adopt this programming model.  The LOPe programming model
explicitly separates numerical complexity from parallel complexity ---
leaving much of the staging and assembling of data to the compiler ---
and is applicable to the solution of stencil-based algorithms that
provide data access patterns that are regular and spatially local.

\section{Programming Model}

The LOPe programming model
restricts the programmer to a local view of the index
space of an array.  Within a LOPe function, only a single array
element (called the local element) is mutable.  In addition, a small
halo region surrounding the local element is visible to the
programmer, but this region is immutable.  Restricting the programmer
to a local index space serves to reduce complexity by separating all
data- and task-decomposition concerns from the implementation of the
element-level array calculations.


LOPe is a domain specific language (DSL) implemented as a small extension to the Fortran 2008
standard.  Fortran was chosen as the base language for LOPe because it provides a rich array-based
syntax.  Although, in principle, the same techniques could be applied to languages such as C or C++.


\subsection{Related work}

LOPe builds upon prior work studying how to map Fortran to accelerator
programming models like OpenCL.  In the ForOpenCL project~\cite{Sottile:2013:FTE:2441516.2441520}
we exploited Fortran's pure and elemental functions to express
data-parallel kernels of an OpenCL-based program.  In practice, array
calculations for a given index $i,j$ will require read-only access to
a local neighborhood of size $[M,N]$ around $i,j$.  LOPe extends this work by introducing
a mechanism for representing these neighborhoods as array declaration
type annotations.

ForOpenCL was based on concepts explored in the ZPL programming language~\cite{chamberlain04zpl} in
which the programmer can define regions and operators that are applied over the index sets
corresponding to the sub-array regions.  This approach is quite powerful for
compilation purposes since it provides a clean decoupling of the operators applied over an array
from the decomposition of that array over a potentially complex distributed memory hierarchy.
However, unlike the ZPL operations on entire sub-arrays, LOPe expresses operations based on
a \emph{single} local array-element.

%
%


\subsection{LOPe Syntax Extensions}

There are only a few syntactic additions required for a LOPe program.
These additions include syntax for describing halo regions and
concurrent procedures.  In code examples that follow,
language additions are highlighted by the usage of capitalization for
keywords that are either new or that acquire new usage.

\subsubsection{Halo regions.}
The principle semantic element of LOPe is the concept of a halo.
A halo is an ``artificial'' or ``virtual'' region surrounding
an array that contains boundary-value information.  Halo (also called
ghost-cell) regions are commonly employed to unify array indexing
schemes in the vicinity of an array boundary so that an array may be
referenced using indices that fall ``outside'' of the logical domain
of the array.  In LOPe, the halo region is given explicit syntax so
that the compiler can exploit this information for purposes of memory
allocation, data replication and thread synchronization.  For example,
a halo region can be declared with a statement of the form,

\begin{verbatim}
  real, allocatable, dimension(:), HALO(1:*:1) :: A
\end{verbatim}
This statement indicates that \texttt{A} is a rank one array, will be
allocated later, and
has a halo region of one element surrounding the array on either side.
The halo notation \texttt{M:*:N} specifies a halo
of \texttt{M} elements to the left, \texttt{N} elements to the
right, and an arbitrary number of ``interior'' array elements.
When used to describe a formal parameter of a
function, such as the type-declaration statement, \texttt{real, HALO(:,:) :: U},
the halo size is inferred by the compiler
from the actual array argument provided at the
calling site of the function.


\subsubsection{Concurrent functions.}

The second keyword employed by LOPe is \texttt{concurrent} which already exists in the form of a
\texttt{do} \texttt{concurrent} loop, whereby the programmer asserts that specific
iterations of the loop body may be executed by the compiler in \emph{any order,} even
\emph{concurrently.}  LOPe allows a function with the attributes \texttt{pure} (assertion of no
side effects) and \texttt{concurrent} (assertion of no dependencies between iterations) to be
called from within a \texttt{do} \texttt{concurrent} loop.  An example of a LOPe function is
shown in Fig. 1 and an example calling this function will be provided later in the text.  One
should imagine that a LOPe function is called \emph{for each} \texttt{i,j} index of the interior of
the array \texttt{U}.  Note that this usage introduces a race condition as new values of
elements of \texttt{U} are created on the left-hand side of the assignment statement that may use
\emph{new or old} values of \texttt{U} on the right-hand side.  LOPe requires the compiler to
guarantee that race conditions won't occur by using, e.g., double-buffering techniques as needed.

\vspace{-.1in}

\begin{figure}
\begin{verbatim}
           pure CONCURRENT subroutine Laplacian(U)}
               real, HALO(:,:) :: U
               U(0,0) =                 U(0,+1)              &
                        +  U(-1,0)  - 3*U(0, 0)  +  U(+1,0)  &
                                    +   U(0,-1)
           end subroutine Laplacian
\end{verbatim}
\vspace{-.1in}
\caption{A LOPe function implementing a Laplacian kernel in two dimensions.}
\end{figure}

\vspace{-.3in}

\subsubsection{LOPe index notation.}

In the \texttt{Laplacian} example the \texttt{U(0,0)} array element is the \emph{local} array
element and only the local element may be modified.  This zero-based indexing for the local-array
element differs from conventional Fortran, where by default, array indices start at 1.  The use of
zero-based indexing gives a clean symmetry for indices on either side of the central element at
zero. The other array elements are in the halo region and are \texttt{U(-1,0)} and
\texttt{U(+1,0)} (left and right of local, respectively) and \texttt{U(0,-1)} and \texttt{U(0,+1)}
(below and above of local).  The geometric positioning of the array elements can be
seen by examining the arrangement of the expressions on the right-hand side of Fig. 1.

\section{Coarray Fortran Extensions}


Consider the \texttt{Laplacian} concurrent function in Fig. 1.  In this section we demonstrate how
this function can be called in the normal context of a program, one that allows full access to all
of the interior elements of the array, as well as array elements within the logically exterior,
halo-boundary region.  Topics highlighted in this section are: 1. distributed memory array
allocation; 2. explicit memory placement; 3. remote memory transfer; and 4. remote execution.  This
description is within the context of extensions to Fortran; as shorthand, these extensions are
referred to as CAFe, for Coarray Fortran extensions.  CAFe is complementary to previous work
extending coarray Fortran\cite{mellor-crummey:2009:caf2,jin:2011:caf2}.

\subsection{Subimages}

We must first introduce the important new concept of a \emph{CAFe subimage}.  Fortran images are a collection of
distributed memory processes that all run the same program (image).  LOPe extends the concept of a
Fortran image by allowing images to be hierarchical.  By this we mean that each image \emph{may}
have a subimage (or subimages), but this subimage is not visible to other regular Fortran images.
Subimages also execute differently than normal images and may execute on different non-homogeneous
hardware, e.g., an attached accelerator device.  Subimages are task based while images all execute
a Single Program but with different (Multiple) Data (SPMD).  A task can be given to a subimage, but
execution on the subimage terminates once the task is finished.  Memory on a subimage is permanent,
however, and must be explicitly allocated and deallocated.

One obtains a subimage by executing the new LOPe function call, \texttt{device = GET\_SUBIMAGE(1)},
where the integer argument represents an attached hardware device (or a separate process).
If the function fails (e.g., the requested device is unavailable) it returns the image number
\texttt{this\_image()} of the process that is executing the current program.  Returning the current
image allows program execution to proceed correctly even if there are no attached devices.

\subsection{CAFe Example}

We start with the declaration of an array with an explicit halo size and with two local
dimensions (rank) and two distributed memory codimensions (corank),
\begin{verbatim}
   real, allocatable, dimension(:,:), codimension[:,:]      &
         HALO(1:*:1,1:*:1) :: U
\end{verbatim}
The corank of the array is chosen to be identical to the rank of the array so that the logical
process topology aligns in a way that allows a natural halo exchange between logically neighboring
processes (this could not occur if corank and rank are not the same).  For example, if the process
location is \texttt{[pcol,prow]}, then the right-hand halo for the local array \texttt{U} can be
obtained by the assignment
\texttt{U(M+1,:) = U(1,:)[pcol+1,prow]} where the size and cosize of
\texttt{U} are given by the allocation statement,
\texttt{allocate(U(0:M+1,0:N+1)[MP,*])}.
This allocation statement specifies (given the one element halo size provided earlier for
\texttt{U}) that the left halo column is \texttt{U(0,:)}, the right column is \texttt{U(M+1,:)},
the bottom row is \texttt{U(:,0)} and the top row is \texttt{U(:,N+1)}, leaving the interior region
\texttt{U(1:M,1:N)}.

In this allocation statement, the total number of process columns $NP$ can be obtained at runtime,
but \emph{may not} be explicitly provided (according to Coarray Fortran (CAF) rules) because the actual number of
participating processes (in Fortran called images) is variable, depending on how many processes are
requested at program startup.  In this discussion, it is \emph{assumed} that there are no holes in
the logical processor topology, thus $MP*NP = P$, where $MP$ is the number of process rows and $P$
is the total number of participating processes (images).

Once a subimage is obtained, memory on the device can be allocated,
\begin{verbatim}
   if (device /= this_image()) then
      allocate(U[device], HALO_SRC=U)   [[device]]
   end if
\end{verbatim}
There are four points to note regarding this memory allocation: 1. Memory is only allocated if a
subimage has been obtained; 2. The location where memory is allocated is denoted by regular coarray
notation \texttt{U[device]}; 3. The allocated size and halo attribute of the new array are obtained
from the previously allocated local array \texttt{U} via the notation \texttt{HALO\_SRC=U} (using
\texttt{HALO\_SRC} will also initially copy \texttt{U} to the subimage); and finally 4. The
allocation itself is \emph{executed} on the subimage device with the notation \texttt{[[device]]}.

Fortran uses square bracket notation, e.g. \texttt{[image]}, to specify on what process the
memory reference is physically located.  Square brackets are a visual clue to the
programmer that the memory reference may be remote and therefore potentially suffer a
performance penalty.  CAFe extends this by employing double-bracket notation to indicate
possibly \emph{remote subimage execution}.

Execution of the \texttt{Laplacian} task is done using the \texttt{do}
\texttt{concurrent} construct:
\begin{verbatim}
   do while (.not. converged)
      do concurrent (i=1:M, j=1:N)   [[device]]
         call Laplacian( U(i,j)[device] )
      end do
      call HALO_TRANSFER(U, BC=CYCLIC)
   end do
\end{verbatim}
There are several points that require highlighting: 1. Iteration occurs over the interior
of the array domain, \texttt{(i=1:M, j=1:N)}; 2. Execution of the loop body occurs on the
specific subimage indicated by \texttt{[[device]]}; 3. Execution of the iterates may occur
in any order, even \emph{concurrently}; 4. The local element of the array (as
defined above in reference to the definition of the concurrent procedure
\texttt{Laplacian}) is given by the indices \texttt{(i,j)}; 5. Location of memory for the
task is to be taken from the subimage as noted by \texttt{[device]}; 6. All threads must finish
execution of the loop body before further execution of the program proceeds; and 7. Transfer of
all requisite halo regions is effected by the call to the new LOPe intrinsic function
\texttt{HALO\_TRANSFER()}.  This function is a synchronization event in that all images must
complete the halo transfer before program execution continues.

Note that a transfer of halo memory is necessary after each completion of the do concurrent loop.
This must be done in order for the halo region of a coarray on a given process to be consistent
with the corresponding interior of the coarray on a logically neighboring process.
Finally, memory for the entire array \texttt{U} can be copied from the
subimage device with the statement,
\texttt{U = U[device]}, 
and memory deallocation (not shown) is similar to memory allocation.

\subsection{Comparison to Coarray Fortran}

LOPe provides a purely \emph{local} viewpoint; the programmer is only provided read and write access
to the local array element and read access to a small halo region surrounding the local element.
There is simply \emph{no} mechanism provided for the programmer to even know \emph{where} the local
element is in the context of the broader array.  On a distributed memory architecture, the halo
elements may not even be physically located on the same processor.  If executed on a cluster
containing hybrid processing elements (e.g. GPUs), the halo elements may be as far as three hops
away: one to get to the host processor and another two to get to memory on the hybrid processor
executing on another distributed memory node.  LOPe provides a complete separation between algorithm
development and memory management (synchronization between memory copies of the same logical
array region covered by halos).  By explicitly describing the existence and size of an array's halo
region, the compiler is provided with enough information to manage most of the hard and detailed
work involved in memory transfer and synchronization.  Additionally, the semantics of the LOPe
execution model remove the possibility of race conditions developing during execution of a
concurrent procedure.

We emphasize some of these advantages by comparing the \texttt{Laplacian} implementation in Fig. 1 with the
implementation of the same algorithm from the original Numrich and Reid
paper\cite{Numrich:1998:CFP:289918.289920} first describing coarrays in Fortran.  We should point
out that this comparison is somewhat unfair, because Numrich and Reid were introducing coarray
notation for transferring memory on distributed memory architectures, not demonstrating how ideally
one should use coarrays within a large application.

However this example serves
to highlight some of the advantages of LOPe and CAFe as introduced above.  Note that in the coarray
example shown below, type declarations have been removed to save space:
{\small \begin{verbatim}
  subroutine Laplace (nrow,ncol,U)
    left = me-1      ! me refers to the current image
    if (me == 1) left = ncol
    right = me + 1
    if (me == ncol) right = 1
    call sync all    ! Wait for left and right
    new_u(1:nrow) = new_u(1:nrow) + u(1:nrow)[left] + u(1:nrow)[right]
    call sync all
    u(1:nrow) = new_u(1:nrow) - 4.0*u(1:nrow)
  end subroutine
\end{verbatim}}


\subsection{LOPe Advantages}

A comparison of Fig. 1 to the CAF example suggests the following advantages:

\begin{itemize}

\item
LOPe requires the implementation of the algorithm to be separate from the call to effect the halo
transfer.  Removing boundary condition specification
from the algorithm allows the boundary conditions to be changed
without changing algorithm code.

\item
LOPe applies the transfer of halo memory across possibly multiple levels of memory with the LOPe
intrinsic \texttt{TRANSFER\_HALO} function.  Thus the LOPe algorithm can be run on a machine with
many interconnected nodes, each containing hybrid processor cores.

\item
Algorithm implementation is separate from user-specified synchronization, e.g., \texttt{call
  sync\_all}.  In LOPe, synchronization is subsumed in the semantics of the \texttt{CONCURRENT}
attribute and the \texttt{TRANSFER\_HALO} function call.

\item
The algorithm implementation is separate from any specification as to where the array
memory is located.  The CAF example explicitly denotes where memory is located with the
\texttt{[left]} and \texttt{[right]} syntax where left and right specify a processor
topology.

\item
The algorithm implementation is separate from any specification as to where the algorithm
is to be executed.  The CAF example explicitly denotes where a statement is to be executed
with the control flow construct \texttt{if (me == 1)}.

\item
The LOPe implementation is easier to understand and frequently follows the mathematical algorithm
directly.  For example, the CAF implementation of Numrich and Reid adds 4 neighbors plus the center
value to make the implementation with direct remote coarray access possible, while the LOPe example
is able to implement the same algorithm with one statement and no intervening synchronization.

\item
The semantics of LOPe makes explicit management of array temporaries (e.g., \texttt{u} and
\texttt{new\_u}) by the programmer unnecessary.
Because in LOPe the
halo region is a language construct, the compiler is better able to manage temporary
buffers than users on the target hardware platform.

\end{itemize}

\subsection{LOPe Constraints}

Constraints provided by the LOPe language extensions allow the compiler to catch several classes of
errors that otherwise would be the programmer's responsibility:

\begin{itemize}

\item
A programmer is not able to store data to the halo region during execution of a LOPe concurrent
function.  Neither are stores to the local element, followed by a read from the halo region of the
same variable allowed.  If these were allowed, one thread could overwrite another threads data at
undefined times.

\item
A programmer can't make indexing errors in a concurrent routine by going out of bounds of the array
plus halo memory.

\item
A programmer is not able to cause race conditions by forgetting to create and use temporary arrays
properly.

\item
A programmer can't make synchronization errors in calls to LOPe functions as synchronization is
implicit in the \texttt{CONCURRENT} attribute.  A thread running a concurrent procedure is provided
with a copy of its local array element plus halo that is consistent with the state of memory
\emph{at the time of invocation of the procedure}.  Stores to an individual thread's local array
element (by that thread) are never visible to other threads.  Normal coarray programs (like MPI)
require explicit synchronization to ensure that processes arrive at the same program location
before a memory read occurs (for example).  LOPE encourages the creation of small functions and
lets the compiler fuse the functions together to improve performance and to provide necessary
synchronization.

\end{itemize}

%
%


\section{LOPe and CAFe Implementation}

This section briefly describes how LOPe extensions to Fortran have been implemented as
source-to-source transformations via rewrite rules (for expressing basic transformations) and
rewriting strategies (for controlling the application of the rewrite rules).  A LOPe file is
transformed to Fortran and OpenCL files through generative programming techniques using
Stratego/XT tools\cite{Bravenboer200852} and is accomplished in three phases: (1) parsing to
produce a LOPe Abstract Syntax Tree (AST) represented in the Annotated Term
Format (ATerm\cite{DBLP:journals/spe/BrandJKO00}); (2) transformations of LOPe AST nodes to Fortran
and C AST nodes; and (3) pretty-printing to the base Fortran and OpenCL languages.

The foundation of LOPe is the syntax definition of the base language expressed in
SDF (Syntax Definition Formalism) as part of the Open Fortran Project (OFP)\cite{OFP:git:url}.
LOPe is defined in a separate SDF module that extends the Fortran 2008 language standard with 15
context-free syntax rules.  Parsing is implemented in Stratego/XT by a scannerless generalized-LR
parser and the conversion of transformed AST nodes to text is accomplished with simple
pretty-printing rules.

\subsection{Transformations for Concurrent Procedures}

A key component of code generation for LOPe is the targeting of a LOPe \texttt{CONCURRENT}
procedure for a particular hardware architecture.  The execution target can be one
of several choices, including serial execution by the current process via inlining
of the function, parallel execution by inlining with OpenMP compiler directives, or parallel
execution by heterogeneous processing elements with a language like OpenCL.

For this work we have developed rewrite rules and strategies in Stratego/XT to rewrite Fortran AST
nodes to C AST nodes (extended with necessary OpenCL keywords).  The C AST ATerms have mostly a
one-to-one correspondence with Fortran terms: a \texttt{CONCURRENT} procedure is transformed to an
OpenCL kernel; Fortran formal parameters are transformed to C parameters (with a direct mapping of
types); and local Fortran variable declarations are rewritten as C declarations.  Similarly,
Fortran executable statements are rewritten as C statements.  The only
minor complication is mapping the LOPe local, array index view to the global C index space.  This
translation is facilitated by a Fortran symbol table that stores array shape and halo size
information.

\subsection{Transformations at the Calling Site}

Transformations of a LOPe procedure call site are more difficult, though technically straight
forward.  The Fortran function call must be transformed to a call to run the OpenCL kernel
(generated as described above).  This is facilitated by use of the ForOpenCL library which
provides Fortran bindings to the OpenCL runtime\cite{Sottile:2013:FTE:2441516.2441520}.  However,
this usage requires the declaration of extra variables, allocation of memory on the OpenCL device
(subimage), transfer of memory, marshalling of kernel arguments, and synchronization.

These transformations are accomplished using several rewrite stages using Stratego/XT strategies:
(1) a symbol table is produced in the first pass to store information related to arrays including,
array shape, halo size, and allocation status;
(2) additional variables are declared that are used to maintain the OpenCL runtime state, including
the OpenCL device, the OpenCL kernel, and OpenCL variables used to marshall kernel arguments; and
(3) all CAFe code related to subimage usage is desugared (lowered) to standard Fortran with calls to the
ForOpenCL library as needed.

Though not yet available, similar rewrite strategies are planned for targeting programming models
other than OpenCL including parallel execution with OpenMP directives.  In addition, simple serial
execution with function inlining (if desired) will be performed by the regular Fortran compiler once
all CAFe and LOPe code has been desugared to standard Fortran.

\section{Conclusions}

Fortran is a general-purpose programming language and as such it provides limited facilities for
expressing concepts useful for stencil operations. For example, halo regions must be expressed in
terms of the existing syntax of the language and there is no way to specify that the ``interior''
of an array is in any way special from an ``outside'' region.  By not providing support for
stencils in the language, the programmer must make specific choices regarding data-access patterns
and the order of operations on the data.  These choices often hide the opportunity for
optimizations by the compiler \cite{Dubey:2014:SSC:2686745.2686756}.  For example, if the compiler
had knowledge of the semantic intent of halo regions, it could reorder operations so that border
regions were computed \emph{before} interior regions, allowing the transfer of data in halo regions
to overlap with computations on the interior.

Just as Coarray Fortran originally extended Fortran to include domain specific knowledge (parallel
computation) replacing MPI library routines\cite{Numrich:1998:CFP:289918.289920}, LOPe seeks to extend Fortran by
providing domain specific knowledge of stencil operations.
Specifically, LOPe and CAFe together provide:
(1) a local view of stencil operations on data that allows a complete separation of the implementation of a stencil
algorithm with data-access patterns;
(2) memory placement via allocation routines that allow the specification of allocation location;
(3) task execution placement with double-bracket syntax specifying which subimage is to execute a particular operation; and
(4) memory exchange via the \texttt{TRANSFER\_HALO} intrinsic procedure.

%
%


\vspace{-.1in}

\subsubsection{Benefits.}
LOPe proposes to formalize the common, halo software pattern in language syntax, thus providing the
compiler with access to halo information in order to spread computation over more hardware
resources, improve performance, and to reduce complexity for the programmer.  Furthermore, LOPe
semantics provide important \emph{language restrictions} that remove the possibility of race conditions
that occur when multiple threads have write access to overlapping data regions.

\vspace{-.1in}

\subsubsection{Limitations.}
LOPe only supports regular structured grids through Fortran multi-dimensional arrays and a
corresponding multi-dimensional processor layout.  It does not allow the composability of stencils
required by non-linear physics operators, nor does it provide automatic support for the storage of
intermediate results resulting from multiple intermediate update steps.  Adaptive Mesh Refinement
(AMR) Shift Calculus provides a generalized abstraction that addresses many of these concerns
(see\cite{Dubey:2014:SSC:2686745.2686756} and references therein).  However, it may be possible to
support AMR in LOPe through locally-structured grid methods based on the work of Berger and
Oliger\cite{colella2007performance}.  In this instance, LOPe could be used to update the regular
array regions in each rectangular patch.

\vspace{.1in}

By implementing LOPe we have demonstrated that LOPe can be used to easily and succinctly code
the stencil algorithms that are common to many areas of science, and furthermore, that LOPe is
suitable for transformation to languages like OpenCL that support heterogeneous computing.  It
remains to history to ascertain if LOPe is sufficiently general purpose to be included in a
general-purpose programming language or if it is better suited to remain as a DSL and to be used as a
special-purpose preprocessing tool.

\bibliographystyle{plain}
\bibliography{local}{}


\subsubsection*{Acknowledgments.}
This work was supported in part by the Department of Energy Office of Science, Advanced Scientific
Computing Research.  The authors would also like to thank Robert Robey and Wayne Weseloh at Los
Alamos National Laborabory for several stimulating conversations with respect to programming models.

\end{document}